\documentclass[apl,aip,english,twocolumn,reprint]{revtex4} \usepackage[T1]{fontenc}
\usepackage[latin1]{inputenc} \usepackage{babel} \usepackage{txfonts}
\usepackage{epsfig,epsf,psfrag} \usepackage{graphicx} \usepackage{epic,eepic}
\usepackage{color,pstcol}\usepackage{pstricks}
 \usepackage{fancyhdr}  \parindent0em

\vspace{-10.0cm} 
\begin{document} 
\title{Persistent currents in absence of magnetic field in graphene nanorings: The ambiguous role of inter valley scattering} 
\author{Colin Benjamin}
\email{cbiop@yahoo.com}  \affiliation{School of Physical Sciences, National Institute of Science education \& Research, Institute of Physics campus, Sachivalaya Marg, Bhubaneswar 751005, India}

\author{A M Jayannavar}
\affiliation{ Institute of Physics, Sachivalaya Marg, Bhubaneswar 751005, India}


\begin{abstract} 
Persistent currents can arise in normal-metal rings due to a magnetic flux threading the ring in equilibrium.  However, can persistent currents arise in absence of magnetic flux in the same normal-metal rings? Yes they can but in a non-equilibrium set-up. This is known as current magnification.  In this work we show that current magnification can be seen in graphene nanorings. Further, graphene can have electrons polarized with a valley quantum number. Electron scattering between valleys can have a non-trivial effect on these persistent currents including inducing a sign change and generating them for parameters where none existed to begin with.
\end{abstract}
\maketitle

A persistent current can flow in a metallic ring threaded by magnetic flux-this was the famous discovery in 1983 by Buttiker, Imry and Landauer\cite{bil} that propelled Mesoscopic physics to the forefront of condensed matter research. The metallic ring is not coupled to any reservoir- a closed system. At zero temperature, persistent current of magnitude  $ev_{F}/L$ where $v_{F}$ is Fermi velocity and $L$ is length of ring, flows in the ring. This effect has been experimentally verified many times, the latest being in 2009\cite{glazman-sci}. This phenomenon is not just restricted to normal metals but has been shown to exist also in carbon nanotube rings\cite{latil} and recently theoretically predicted to exist in graphene rings\cite{peeters} and topological insulator rings\cite{michetti} too. The reason for the existence of a persistent current is that a magnetic flux breaks the time reversal symmetry between left and right moving charge carriers enabling a net current to flow in the ring. It has been predicted that a persistent current  can be generated in absence of any magnetic flux via transport currents in a non-equilibrium setup albeit in the linear response regime\cite{amj}. However, to see this effect, i.e., persistent currents in absence of magnetic flux we have to be in an open system- the ring has to be coupled to electron reservoirs so that electronic transport is enabled through the ring. This phenomenon is called current magnification. This effect has no classical analog. The importance of observing current magnification in graphene has to be seen in the context of experimental verification of this noteworthy effect, since till date it hasn't been observed experimentally although manifold theoretical proposals exist. Among the notable theoretical proposals include current magnification effect with spin currents\cite{choi}, with thermo-electric currents\cite{moskalets} and heat currents\cite{marathe}. In the literature effect of current magnification is also termed by some as giant persistent current\cite{ye}.

In this work we predict that current magnification can also occur in graphene and increasing intervalley scattering far from having a detrimental effect can actually generate these persistent currents  for parameter values where in absence of intervalley scattering there were none. We also find that this type of scattering can change the sign of these currents. Now what is intervalley scattering? In graphene as is well known electrons are not only distinguished by their spin but by a index called pseudospin (or, valley). This pseudospin is nothing but the fact that in the Energy-momentum diagram, graphene does not have a unique Fermi point but two Fermi points-often described as $k+$ and $k-$. Fermi points are points in the diagram where conduction and valence bands touch. Thus electrons in graphene can have momenta (or valley index) $k+$ or $k-$. Generally scattering between these valleys is non-existent unless there are some lattice inhomogenities (at the atomic scale) which will mix the states from different valleys\cite{kats-book}.  Inter valley scattering when it exists can have dramatic consequences for electron transport in graphene. It has been predicted theoretically that weak localization shouldn't be observed in graphene. Rather, weak antilocalization should be. When intervalley scattering is absent it is weak anti localization that  is seen. Weak localization refers to the fact that resistance in the quantum limit is slightly larger than the classical limit in normal metals and semiconductors\cite{datta}. The reason being quantum interference effects lead to coherent back scattering. The consequence of weak antilocalization in graphene is that in the quantum limit resistance is slightly lower than that in classical limit. This is because of the presence or absence of intervalley scattering. Strong intervalley scattering leads to conventional weak localization in graphene while absence of intervalley scattering is the reason for weak anti-localization.  In this work we show that similar to the case of localization discussed above intervalley scattering leads to anomalous results.

 Graphene is a monatomic layer of graphite with a honeycomb lattice
structure\cite{graphene-rmp} that can be split into two triangular sublattices $A$ and $B$. The electronic properties of  graphene are effectively described by the Dirac
equation\cite{graphene-sudbo}. The presence of isolated Fermi points, $k_+$ and $k_-$, in its spectrum, gives rise to two distinctive valleys. These two Fermi points are ofcourse inter related $k_{\pm}=\pm k$. Further, since electrons in graphene can occupy different valleys\cite{heikkila} and transport can be with or without inter valley scattering, graphene gives us a tool to see how robust is current magnification to this unique form of  scattering.
\begin{figure}[h]
 \centerline{\includegraphics[width=8cm,height=8cm]{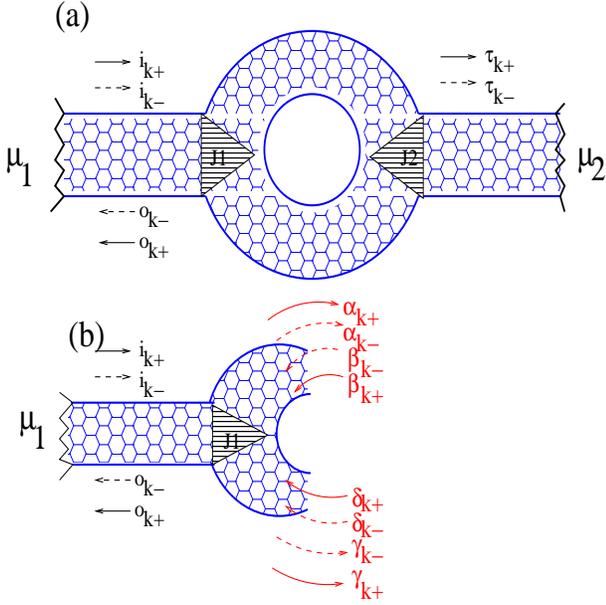}}
\caption{(a). The model system. A circular monolayer graphene ring is connected via graphene leads to  two reservoirs at chemical potentials $\mu_1$ and $\mu_2$.  (b) The waves incident and outgoing from  left coupler $J_1$. } \label{scheme}
\end{figure}

In our model system depicted in Fig.\ref{scheme}, a graphene nanoring is coupled to electron reservoirs kept at different chemical potentials. We use the scattering matrix(S-Matrix) system used earlier in normal metallic systems\cite{butipra} and generalized for graphene\cite{zannen}. Electrons are incident from the left of the ring since $\mu_{1} > \mu_{2}$. Electron transport in graphene has been exhaustively dealt with in Ref. [\onlinecite{heikkila}]. Graphene being a 2D honeycomb lattice structure. Its unit cell has two atoms and the brillouin zone of graphene has two Fermi points where conduction and valence bands touch. Near these valleys the energy band structure of graphene  is linear implying Dirac behavior rather than Schrodinger.  Since there are two valleys one  can have electrons polarized in either one of them or both. Further electrons polarized in one valley can maintain their polarization throughout or can be scattered from one to the other. When we have maximal intervalley scattering, i.e. the weak localization limit, we have one type of situation. However, when intervalley scattering is absent, the weak anti-localization limit we get different results. We take all these factors into consideration while analyzing electron transport through the ring.

\begin{figure}[h]
 \centerline{ \includegraphics[width=8cm,height=5cm]{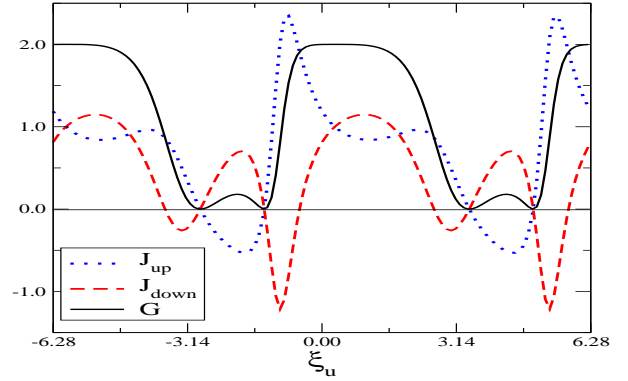}}
\caption{The current densities and conductance as a function of the scattering phase in upper branch $\xi_{u}$ for $\gamma=0.5, \xi_{d}=0.3$.}  \label{fig:G-j1-j2}
\end{figure}

The electronic transport through our system can be easily analyzed via taking recourse to S-Matrix theory\cite{datta}. Conductance through normal metal nanoring systems have been very well analyzed in Refs.\cite{butipra,gefenprl}. This S-Matrix analysis has been extended to conductance through graphene nanorings\cite{zannen}. In this work we follow this prescription\cite{zannen} but adapted to our system. As mentioned earlier electrons in graphene can be with either $k+$ or $k-$ valley momenta. We assume that electrons with momenta $k+$ or $k-$ are incident from the left reservoir $\mu_1$. The S-Matrix for left coupler $J1$ yields the outgoing amplitudes $(o_{k+},o_{k-},\alpha_{k+},\alpha_{k-},\gamma_{k+},\gamma_{k-})$ in terms of the incident amplitudes $(i_{k+},i_{k-},\beta_{k+},\beta_{k-},\delta_{k+},\delta_{k-})$ and for right coupler $J2$ yields the outgoing amplitudes $(\tau_{k+},\tau_{k-},\alpha^{'}_{k+},\alpha^{'}_{k-},\gamma^{'}_{k+},\gamma^{'}_{k-})$ in terms of the incident amplitudes $(i^{'}_{k+},i^{'}_{k-},\beta^{'}_{k+},\beta^{'}_{k-},\delta^{'}_{k+},\delta^{'}_{k-})$. Since nothing is incident from right $i^{'}_{k+}=i^{'}_{k-}=0$ further we assume incident wave amplitudes from left $i_{k+}=i_{k-}=1$.
At the junction $J1$ and $J2$ the S-Matrices take the form-
\begin{equation}
 S_{J1}=\left(\begin{array}{ccc} -(a+b)\bf{I}      & \sqrt\epsilon\bf{I}&\sqrt\epsilon\bf{I}\\ \sqrt\epsilon\bf{I}& a  \bf{I}          &b   \bf{I}         \\ \sqrt\epsilon\bf{I}& b    \bf{I}        &a\bf{I} \end{array} \right)
 \end{equation}
 with $a=\frac{1}{2}(\sqrt{(1-2\epsilon)} -1), b=\frac{1}{2}(\sqrt{(1-2\epsilon)} +1)$, {\bf I} being the identity matrix. $\epsilon$ plays the role of a coupler with maximum coupling $\epsilon=\frac{1}{2}$ while for $\epsilon=0$, the coupler completely disconnects the loop from the lead.

 The electrons scattered from either couplers are related to each other by  the S-matrix  of the upper branch by-

 \begin{equation} \left(\begin{array}{c}\beta_{k+}\\ \beta_{k-}\\ \beta^{'}_{k+}\\ \beta^{'}_{k-}\end{array}\right) =e^{i\xi_{u}} \left(\begin{array}{cccc} 0     & 0&t &a\\ 0 & 0 &a&t\\ t&a&0&0\\ a&t&0&0 \end{array} \right)\left(\begin{array}{c}\alpha_{k+}\\ \alpha_{k-}\\ \alpha^{'}_{k+}\\ \alpha^{'}_{k-}\end{array}\right)
 \end{equation}
with $t=\sqrt{1-\gamma^{2}}, a=i\gamma, \gamma \in [0,1], \xi_{u}$ is an overall effective phase which denotes magnitude of inter-valley scattering in the upper branch\cite{zannen}. The probability of transmission in same valley is $|t|^{2}$ and the probability of transmission with scattering to opposite valley is $|a|^{2}=\gamma^{2}$. $\gamma=0$ correspond to infinite intervalley scattering time while $\gamma=1$ denotes zero intervalley scattering time, i.e., electron  in $k_{+}$ valley will be immediately scattered to $k_{-}$ valley.

Exactly the same form of the S-Matrix applies to lower branch but with $\xi_{u}$ replaced by $\xi_{d}$:
 \begin{equation} \left(\begin{array}{c}\delta_{k+}\\ \delta_{k-}\\ \delta^{'}_{k+}\\ \delta^{'}_{k-}\end{array}\right) =e^{i\xi_{d}} \left(\begin{array}{cccc} 0     & 0&t &a\\ 0 & 0 &a&t\\ t&a&0&0\\ a&t&0&0 \end{array} \right)\left(\begin{array}{c}\gamma_{k+}\\ \gamma_{k-}\\ \gamma^{'}_{k+}\\ \gamma^{'}_{k-}\end{array}\right)
 \end{equation}

Combining the S-Matrices of the junctions and upper and lower branch we can calculate the scattering amplitudes in the different parts of the ring. We will have different amplitudes for each valley. There are twelve scattering probabilities which we will get after solving the system. The conductance for each valley which in this two terminal system is just the transmission probability $|\tau_{k+}|^2$, and $|\tau_{k-}|^2$, the  current amplitudes for each valley in either branch of the  ring and the probability for reflection. The current densities for each valley in either branch is the difference between the squared modulus of the amplitudes for incoming and outgoing amplitudes, i.e., for upper branch $Iu_{k+}=|\beta_{k+}|^2-|\alpha_{k+}|^2$ and $Iu_{k-}=|\beta_{k-}|^2-|\alpha_{k-}|^2$. The total conductance and current densities is the sum of individual valley conductances and individual valley current densities which implies $G=|\tau_{k+}|^2+|\tau_{k-}|^2$ and $J_{u}=Iu_{k+} + Iu_{k-}$, while $J_{d}=Id_{k+} + Id_{k-}$.

In the figures below, we take the junction coupling strength $\epsilon=0.5$, this implies the junction does not back reflect any incident electrons and  we plot the current densities and conductance for different parameters of intervalley scattering parameter $\gamma$, the overall scattering phase for upper branch $\xi_{u}$ and for lower branch $\xi_{d}$.
\begin{figure}
\vskip 0.25in
\centerline{
\includegraphics[width=8cm,height=5cm]{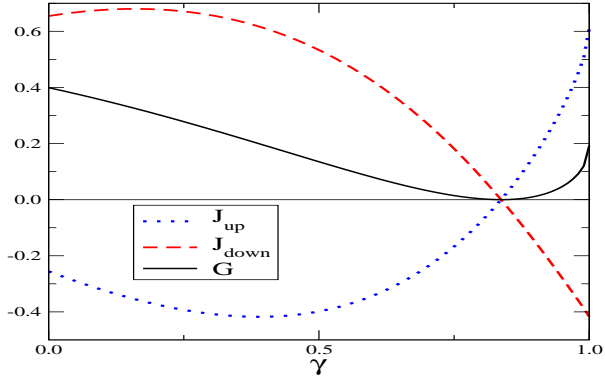}} \caption{The current densities and conductance as a function of the inter valley scattering parameter $\gamma$.}   \label{fig:G-j1-j2-gam}
\end{figure}

\begin{figure}
 \vskip 0.25in
\centerline{
\includegraphics[width=8cm,height=5cm]{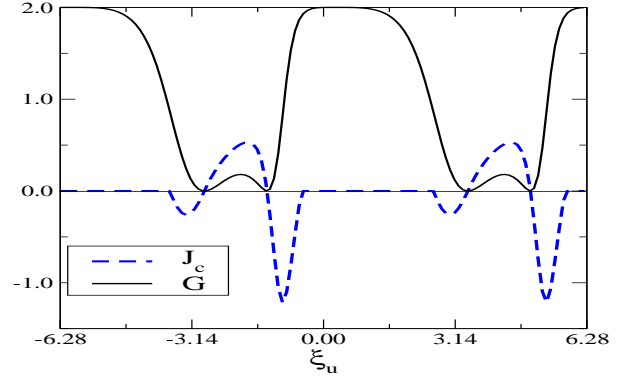}} \caption{The persistent current density and conductance as function of the scattering phase in upper branch $\xi_{u}$ for $\gamma=0.5, \xi_{d}=0.3$.} \label{fig:jc-gam}
\end{figure}

\begin{figure}[h]
\vskip 0.25in \centerline{\includegraphics[width=8cm,height=5cm]{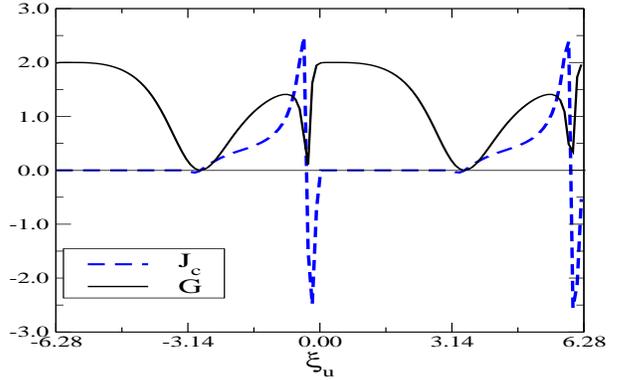}}
 \caption{Persistent current and conductance as function of the scattering phase in upper branch $\xi_{u}$ for $\gamma=0$.} \label{fig:jc-gam0}
 \end{figure}
In Fig.\ref{fig:G-j1-j2} we plot the current densities in upper branch and lower branch alongwith the total conductance for $\gamma=0.5$ and $\xi_{d}=0.3$ as function of $\xi_{u}$. We see that for a range of values  the current density in upper or lower branch may be larger than the total transport current density (or, the transmission probability which in the Landauer-Buttiker picture is the conductance\cite{datta}). This clearly seems to be an apparent violation of Kirchoff's law of current conservation. However it is not so since the current density in the other branch is negative as is evident so as not to violate the Kirchoff's law. However, what does this negative current density mean? This negative current density is interpreted as a persistent current which flows continually in the loop against the direction of the transport current. The sign of the persistent current is taken as follows when the current density in lower branch is negative then direction of persistent current is  positive and if current density in upper branch is negative then persistent current direction is  negative\cite{pareek,colin}.  Here these persistent currents are not produced via magnetic flux but via transport currents in a non-equilibrium set-up, as long as a voltage bias remains persistent current follows.

In Fig. \ref{fig:G-j1-j2-gam} the current densities and conductance are plotted against the inter valley scattering parameter $\gamma$. We see that for the parameters chosen the current magnification effect is observed throughout the range of $\gamma$ implying that  current magnification effect is quite resilient to intervalley scattering which has a non-trivial effect on transport. Since as has been seen increasing intervalley scattering is the reason for weak anti-localization changing to weak localization. 

In Figs.\ref{fig:jc-gam} and \ref{fig:jc-gam0} the persistent current density and conductance are plotted as function of  the scattering phase shift $\xi_u$ encountered in the upper branch for $\gamma=0.5$ and $\gamma=0$. persistent currents arise near the zeros of the conductance. Similar to those seen in normal metals rings current magnification effect in graphene is seen at the poles of the transmission amplitude.

\begin{figure}[h]
\vskip 0.25in \centerline{\includegraphics[width=8cm,height=5cm]{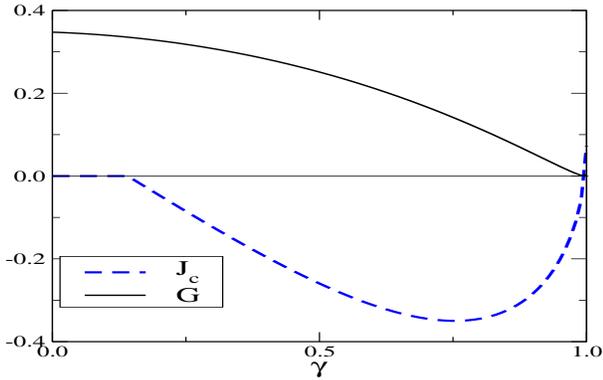}}
\caption{Persistent current and conductance as a function of $\gamma$ with $\xi_{u}=3.0,\xi_{d}=0.3$. At $\gamma\rightarrow 0$- the weak antilocalization limit, there is no current magnification while for $0.2 < \gamma < 1.0$ a finite persistent current flows and near the weak localization limit $\gamma \sim 1.0$ the persistent current density even changes sign}
\label{fig:jc-gam-xi-u-3}
\end{figure}

\begin{figure}[h]
\vskip 0.25in
\centerline{\includegraphics[width=8cm,height=5cm]{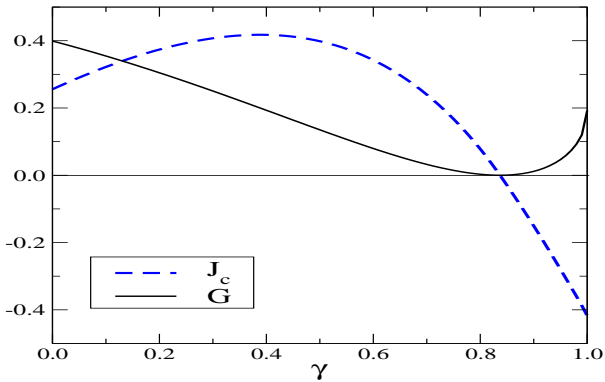}}
\caption{Persistent current and conductance as a function of $\gamma$ with $\xi_{u}=4.0,\xi_{d}=0.3$. In this case contrary to Fig.~\ref{fig:jc-gam-xi-u-3} current magnification effect exists in the weak antilocalization limit. }
\label{fig:jc-gam-xi-u4}
\end{figure}

\begin{figure}[h]
\vskip 0.25in
\centerline{\includegraphics[width=8cm,height=5cm]{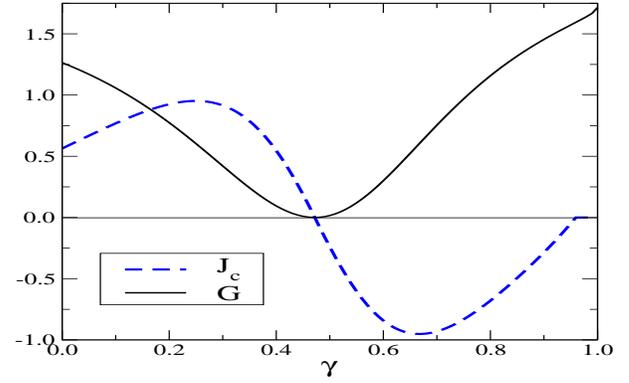}}
\caption{Persistent current and conductance as a function of $\gamma$ with $\xi_{u}=5.0,\xi_{d}=0.3$. Here current magnification disappears in the weak localization limit.}
\label{fig:jc-gam-xi-u5}
\end{figure}
We analyze the behavior of the persistent current density as function of $\gamma$ for $\xi_{d}=0.3$ and for different values of $\xi_{u}=3,4$ and $5$ in Figs.\ref{fig:jc-gam-xi-u-3},\ref{fig:jc-gam-xi-u4} and \ref{fig:jc-gam-xi-u5}. One thing which is quite evident is that the persistent current density can increase with increase in intervalley scattering which is quite counterintuitive. In Fig.\ref{fig:jc-gam-xi-u5} it is very evident that at zero $\gamma$ there is no persistent current which is finite at  $\gamma \gg 0$, further for $\gamma \gg 0.5$ there is a sign change in peristent currents, a paramagnetic persistent current can change to a diamagnetic one this can also be seen in Fig.\ref{fig:jc-gam-xi-u4} for $\gamma$ values greater than $0.85$.

Now we analyze the reasons for current magnification effect. The reason is the phase difference between the two branches $\xi_{u}\neq\xi_{d}$. If it were not so there wouldn't be any current magnification. Classically our system mirrors the case of two parallel resistors. We know that currents flowing in case of parallel resistors are never negative. However, when we quantum mechanically add two resistors in parallel the overall resistance does not follow Ohms law $1/R\neq 1/R_{1}+1/R_{2}$, since there is included a term in the total resistance which depends on the phase difference between two resistors\cite{datta,gefenprl}. When the same calculation is done for individual currents one can come across situations where current in either branch may turn negative giving rise to current magnification.

Finally we analyze the effects of intervalley scattering. As has been seen before, intervalley scattering is the reason for weak localization in graphene. In our work we see that current magnification can be generated because of intervalley scattering (as in Fig.~\ref{fig:jc-gam-xi-u-3}) and also it can be destroyed by maximal intervalley scattering (as in Fig.~\ref{fig:jc-gam-xi-u-3}). One thing is certain that intervalley scattering does change the sign of persistent currents in graphene.

The experimental realization of this structure is not at all difficult, graphene nanorings have been designed in many contexts\cite{ihn}. We have a transport current which flows through the nanoring, this  leads to a persistent current in the ring. Amperes law states that magnetic moment in a ring is product of current and the area enclosed by the current flowing in the ring. Existence of a persistent current in the ring will lead to a magnetic moment in the ring. This magnetic moment can be measured. The scattering phase shifts $\xi_u$ and $\xi_d$ are dependent on valley momentum which can be easily manipulated via a gate voltage making the current magnification prediction that much more testable.

CB would like to thank DST Nanomission for the grant No. SR/NM/NS-1101/2011 under which a substantial portion of the work was completed.
AMJ would like to thank Dept. of Science \& Technology, Govt. of India for financial support.

\end{document}